\documentclass[apj]{emulateapj}

\shorttitle{Redshift Evolution of AGN bias}
\shortauthors{Allevato et al.}

\begin{document}

\slugcomment{Accepted for publication in The Astrophysical Journal}

\title{Clustering of moderate luminosity X-ray selected type 1 and type 2 AGN at z$\sim$3}

\author{V. Allevato\altaffilmark{1}, 
A. Finoguenov\altaffilmark{1,2}, 
F. Civano\altaffilmark{3,4}
N. Cappelluti\altaffilmark{5,2}, 
F. Shankar\altaffilmark{6,7}
T. Miyaji\altaffilmark{8,9}, 
G. Hasinger\altaffilmark{10}, 
R. Gilli\altaffilmark{5}, 
G. Zamorani\altaffilmark{5}, 
G. Lanzuisi\altaffilmark{11}, 
M. Salvato\altaffilmark{12}, 
M. Elvis\altaffilmark{4},
A. Comastri \altaffilmark{5} \&
J. Silverman\altaffilmark{13}}

\altaffiltext{1}{Department of Physics, University of Helsinki, Gustaf H\"allstr\"omin katu 2a, FI-00014 Helsinki, Finland}
\altaffiltext{2}{University of Maryland, Baltimore County, 1000 Hilltop Circle, Baltimore, MD 21250, USA}
\altaffiltext{3}{Department of Physics and Astronomy, Dartmouth College, 6127 Wilder Laboratory, Hanover, NH 03755}
\altaffiltext{4}{Harvard Smithsonian Center for Astrophysics, 60 Garden Street, Cambridge, MA 02138, USA}
\altaffiltext{5}{INAF-Osservatorio Astronomico di Bologna, Via Ranzani 1, 40127 Bologna, Italy}
\altaffiltext{6}{Department of Physics and Astronomy, University of Southampton, Highfield, SO17 1BJ}
\altaffiltext{7}{GEPI, Observatoire de Paris, CNRS, Univ. Paris Diderot, 5 Place Jules Janssen, F-92195 Meudon, France}
\altaffiltext{8}{Instituto de Astronomia, Universidad Nacional Autonoma de Mexico, Ensenada, Mexico (mailing adress: PO Box 439027, San Ysidro, CA, 92143-9024, USA)}
\altaffiltext{9}{Center for Astrophysics and Space Sciences, University of California at San Diego, Code 0424, 9500 Gilman Drive, La Jolla, CA 92093, USA}
\altaffiltext{10}{Institute for Astronomy, University of Hawaii, 2680 Wood- lawn Drive, Honolulu, HI 96822, USA}
\altaffiltext{11}{National Observatory of Athens I.Metaxa \& Vas. Pavlou St.
GR-15236 Penteli, GREECE}
\altaffiltext{12}{Max-Planck-Institute f\"ur Extraterrestrische Physik, Giessenbachstrasse 1, D-85748 Garching, Germany}
\altaffiltext{13}{Institute for the Physics and Mathematics of the Universe, The University of Tokyo, 5-1-5 Kashiwanoha, Kashiwashi, Chiba 277-8583, Japan}

\begin{abstract}

We investigate, for the first time at z$\sim$3, the clustering properties
of 189 Type 1 and 157 Type 2 X-ray active galactic nuclei 
(AGN) of moderate luminosity ($\langle L_{bol} \rangle$ = 10$^{45.3}$ erg s$^{-1}$), 
with photometric or spectroscopic
redshifts in the range 2.2$<$z$<$6.8. These samples are based on 
\textit{Chandra} and \textit{XMM-Newton} 
data in COSMOS. 
We find that Type 1 and Type 2 COSMOS AGN at z$\sim$3 inhabit DMHs with typical 
mass of logM$_h$ = 12.84$^{+0.10}_{-0.11}$ and 
11.73$^{+0.39}_{-0.45}$ h$^{-1}$M$_{\odot}$, respectively. This result requires a 
drop in the halo masses of Type 1 and 2 COSMOS AGN 
at z$\sim$3 compared to z$\lesssim$2 
XMM COSMOS AGN with similar luminosities. Additionally, we infer that 
unobscured COSMOS AGN at z$\sim$3 reside in 10 times more massive halos
compared to obscured COSMOS AGN, at 2.6$\sigma$ level. This result extends 
to z$\sim$3 that found in COSMOS at z$\lesssim$2, and rules out the picture in 
which obscuration is purely an orientation effect.
A model which assumes that the AGN activity is triggered
by major mergers is quite successful in predicting both the low halo mass
of COSMOS AGN and the typical mass of luminous SDSS quasars at z $\sim$ 3,
with the latter inhabiting more massive halos respect to
moderate luminosity AGN.
Alternatively we can argue, at least for Type 1 COSMOS AGN, that they are possibly representative of an early phase of fast (i.e. Eddington limited) BH growth
induced by cosmic cold flows or disk instabilities.
Given the moderate luminosity, these new fast 
growing BHs have  
masses of $\sim 10^{7-8}$ M$_{\odot}$ at z$\sim$3 which might evolve into 
$\sim 10^{8.5-9}$ M$_{\odot}$ mass BHs at z=0.
Following our clustering measurements, we argue that 
this fast BH growth at z$\sim$3 in AGN with moderate luminosity occurs in DMHs with typical 
mass of $\sim$ 6$\times 10^{12}$ h$^{-1}$M$_{\odot}$.

\end{abstract}


\section{Introduction}
\label{sec:intro}

The connection between black holes (BHs) and their 
host dark matter halos (DMHs) has been mainly 
studied via clustering measurements of active galactic nuclei (AGN).
Under an assumed cosmology, the AGN bias 
(i.e. the square root of the relative amplitude of AGN 
clustering to that of dark matter, e.g., Kaiser 1984) 
can be inferred and linked to the typical mass of 
AGN hosting DMHs (e.g., Jing 1998; Sheth \& Tormen 1999, 
Sheth et al. 2001, Tinker et al. 2005, 2010). This
provides information about 
galaxy/AGN co-evolution and 
the mechanisms that trigger the AGN activity.

The clustering properties of 
thousands of broad-line luminous quasars with 
typical L$_{bol} \gtrsim 10^{46}$ erg s$^{-1}$,
have been studied in different large area 
optical surveys, such as 2QZ (e.g. Croom et al. 2005; 
da Angela et al. 2005; Porciani \& Norberg 2006), SDSS 
(e.g., Shen et al. 2009; Ross et al. 2009)
and 2SLAQ (Croom et al. 2009; da Angela et al. 2008).
All these studies suggest the common picture that
luminous optically selected quasars are hosted by halos of roughly 
constant mass, a few times 10$^{12}$ M$_{\odot}$ h$^{-1}$, out to z$\sim$3-4.
This lack of variation in halo mass implies 
that the bias factor is an increasing function of 
redshift, since the DM is more weakly clustered earlier in cosmic time.

In addition, quasar clustering measurements
have also facilitated several theoretical investigations on the
cosmic evolution of BHs within the hierarchical structure
formation paradigm (e.g., Hopkins et al. 2007; Shankar et al.
2008a, 2008b; White et al. 2008; Croton 2008; Wyithe \& Loeb
2008). Interestingly, models of major mergers between 
gas-rich galaxies appear to naturally produce the evolution of 
the quasar large-scale bias as a function of luminosity and redshift 
(Hopkins et al. 2007, 2008; Shen 2009; Shankar et al. 2009, 
2010; Bonoli et al. 2009). This supports the scenario in which 
major mergers dominate the luminous quasar population
(Scannapieco et al. 2004, Shankar 2010, Neistein \& Netzer 2013, 
Treister et al. 2012).

X-ray detection is generally recognized as a 
more robust way to obtain a uniformly selected AGN sample
with lower luminosities (L$_{bol} \sim 10^{44-46}$ erg s$^{-1}$)
and with a significant fraction of
obscured sources with respect to optical surveys. 
This means that while deep X-ray AGN samples are from 
square-degree area surveys, sampling 
moderate luminosity AGN, optical quasars 
are from thousands of square degree surveys, 
sampling rare and high luminosity AGN events.
Thanks to \textit{Chandra} and \textit{XMM-Newton} surveys, 
large samples of X-ray AGN are available
and clustering measurements of moderate luminosity
AGN are now possible with a precision comparable to that achievable 
with quasar redshift surveys.

Measurements of the spatial clustering of X-ray AGN show that they are located in galaxy 
group-sized DMHs with logM$_h$ = 13-13.5 h$^{-1}$M$_{\odot}$
at low ($\sim$0.1) and high ($\sim$ 1-2) redshift 
(e.g. Hickox et al. 2009, 
Cappelluti et al. 2010, Allevato et al. 2011,
Krumpe et al. 2010, 2012,
Mountrichas et al. 2012, Koutoulidis et al. 2013). 

The fact that DMH masses of this class of moderate luminosity 
AGN are estimated to be, on average, 5-10 times larger than 
those of luminous quasars, has been interpreted as evidence against 
cold gas accretion via major mergers in those systems 
(e.g. Allevato et al. 2011; Mountrichas \& Georgakakis 2012).
Additionally, it has been explained
as support for multiple modes of BH accretion 
(cold versus hot accretion mode, e.g. Fanidakis et al. 2013).
However, this difference, which may not be present 
at z$<$0.7 (Krumpe et al. 2012) does 
not yet have a good explanation.

On the other hand, several works on the morphology of the 
AGN host galaxies suggest that, even at moderate luminosities, a large 
fraction of AGN is not associated with morphologically disturbed galaxies.
This trend has been observed both at low (z$\sim$1, e.g. Georgakakis et 
al. 2009, Cisternas et al. 2011)
and high (z$\sim$2, e.g. Schawinski et al. 2011, Rosario et al. 2011, 
Kocevski et al. 2012)
redshift.


Despite the power of clustering measurements in understanding 
AGN population, little is known about the clustering of obscured AGN.
These sources, based on the results from deep X-ray surveys 
(e.g. Brandt \& Hasinger 2005; Tozzi et al. 2006) and 
X-ray background synthesis models (e.g. Civano et al. 2005, Gilli et al. 2007), 
are the most abundant AGN population in the Universe. Additionally, they are 
expected to dominate the history of accretion onto SMBHs (e.g. Fabian \& Iwasawa 1999).
A basic prediction of orientation-driven AGN unification 
models is that the clustering strength should be similar 
for obscured (narrow-line or Type 2) and 
unobscured (broad-line or Type 1) AGN. 
By contrast, in the AGN evolutionary scenario, obscured quasars 
may represent an early evolutionary phase after a major 
merger event, when 
the growing BHs can not produce a high enough 
accretion luminosity to expel the surrounding material 
(e.g., Hopkins et al. 2008; King 2010). Following this argument, 
the luminous quasar phase might probably correspond
to the end of an obscured phase.
On the other hand, if the AGN activity
is triggered by sporadic gas inflow, not by 
major mergers, then obscured and unobscured
AGN might be two stages that may occur several times
along the galaxy lifetime. The different durations of these 
two stages and their relation to the environment 
may produce different clustering properties between 
obscured and unobscured AGN (Hickox et al. 2011).

Some studies of optically selected quasars 
confirm that low-redshift narrow-line AGN 
are not strongly clustered and are hosted in galaxies that 
do not differ significantly from typical non-AGN galaxies 
(e.g., Wake et al. 2004, Magliocchetti et al. 2004, 
Mandelbaum et al. 2009; Li et al. 2006). 
Hickox et al. (2011), analysing a sample of 
806 Spitzer mid-IR-selected quasars at 0.7 $<$ z $<$ 1.8 
in the Bootes field, find marginal ($<$ 2$\sigma$) 
evidence that obscured quasars have a larger bias and 
populate more massive DMHs than unobscured quasars.
Recently, Donoso et al. (2013)
using mid-IR-WISE selected AGN candidates at z$\sim$1.1,
infer that red AGN (i.e obscured sources) are hosted 
by massive DMHs of logM$_h$ $\sim$ 
13.5 h$^{-1}$M$_{\odot}$. This value is well above the  
halo mass of logM$_h$ $\sim$ 12.4  h$^{-1}$M$_{\odot}$
that harbour blue AGN (unobscured sources). 

On the contrary, Krumpe et al. (2012)
find no significant difference in the clustering of 
X-ray narrow-line and broad-line RASS AGN
at 0.07$<$z$<$0.5.
A larger clustering amplitude for Type 1 
with respect to Type 2 AGN, has been observed in 
the Swift- BAT all sky survey at z$\sim$0 (Cappelluti et al. 2010).
The redshift evolution of the bias of moderate luminosity X-ray  
AGN have been investigated in Allevato et al. (2011)
by using XMM COSMOS data.
They find that the bias increases with redshift tracing 
a constant halo mass typical of galaxy groups ($\sim 10^{13}$ h$^{-1}$M$_{\odot}$)
up to z$\sim$2. Additionally, their results indicate that
obscured XMM COSMOS AGN inhabit slightly (2.3$\sigma$) less 
massive halos than unobscured sources.

The clustering of AGN at z $>$ 2
is still poorly investigated. 
At high redshifts galaxies and AGN are thought 
to form in rare peaks of the 
density field and then to be strongly biased relative 
to the DM (Kaiser 1984; Bardeen et al. 1986).
The clustering of z $>$ 2.9 SDSS quasars (Shen et al. 2007, 2009)
indicates (with large uncertainties on the bias) that luminous quasars 
reside in massive halos with mass few times $10^{13}$ h$^{-1}$M$_{\odot}$.
Following Shankar et al. (2010), these clustering measurements  
require a high duty cycle
(i.e. the probability for an AGN of being active at a given time)
for massive BHs ($>10^{9}M_{\odot}$) in luminous quasars 
(L$_{bol}>10^{46}$ erg s$^{-1}$).
The clustering signal measured 
by Shen et al. (2009) at z=3.2 has also been interpreted with the 
halo occupation distribution (HOD) by Richardson et al. (2012).
Given the large uncertainty of the signal at z=3.2,
especially at small scales, they only
infer the mass of central halos hosting quasars 
(M$_{cen}$=14.1$^{+5.8}_{-6.9}$ 
$\times10^{12}$ h$^{-1}$M$_{\odot}$).

The clustering of moderate luminosity X-ray AGN
at z $\geqslant$ 2 is indeed largely unexplored.
The only attempt of measuring the clustering properties of X-ray AGN at 
z=3 is presented in Francke et al. (2008). They estimate the
correlation function of a small sample of 
X-ray AGN with L$_{bol} \sim 10^{44.8}$ erg s$^{-1}$, 
in the Extended Chandra Deep Field South (ECDFS).
They find indications that X-ray ECDFS AGN reside in DMHs with 
minimum mass of logM$_{min}$ = 12.6$^{+0.5}_{-0.8}$ h$^{-1}$ M$_{\odot}$.
Unfortunately, because of the small number of sources,
the bias factor has a very large uncertainty.

In this paper we use a larger sample of X-ray selected AGN
with L$_{bol} \sim 10^{45.3}$ erg s$^{-1}$,
based on \textit{Chandra} and \textit{XMM-Newton} data in the COSMOS field,
at 2.2$<$z$<$6.8. The purpose is to measure the 
clustering amplitude and the typical hosting halo mass of moderate 
luminosity AGN at z $\sim$ 3. Additionally, 
we focus on the measurements of the large-scale bias of Type 1 and Type 2 COSMOS AGN
at z $\geqslant$ 2.2. This redshift range has never been explored 
before for the clustering of moderate luminosity obscured and unobscured sources.
Throughout the paper, all distances are 
measured in comoving coordinates and are given in
units of Mpc $h^{-1}$, where $h=H_0/100$km/s. 
We use a $\Lambda$CDM cosmology with
$\Omega_M=0.3$, $\Omega_\Lambda=0.7$, 
$\Omega_b=0.045$, $\sigma_8=0.8$.
The symbol $log$ signifies a base-10 logarithm.

\begin{figure*}
\plotone{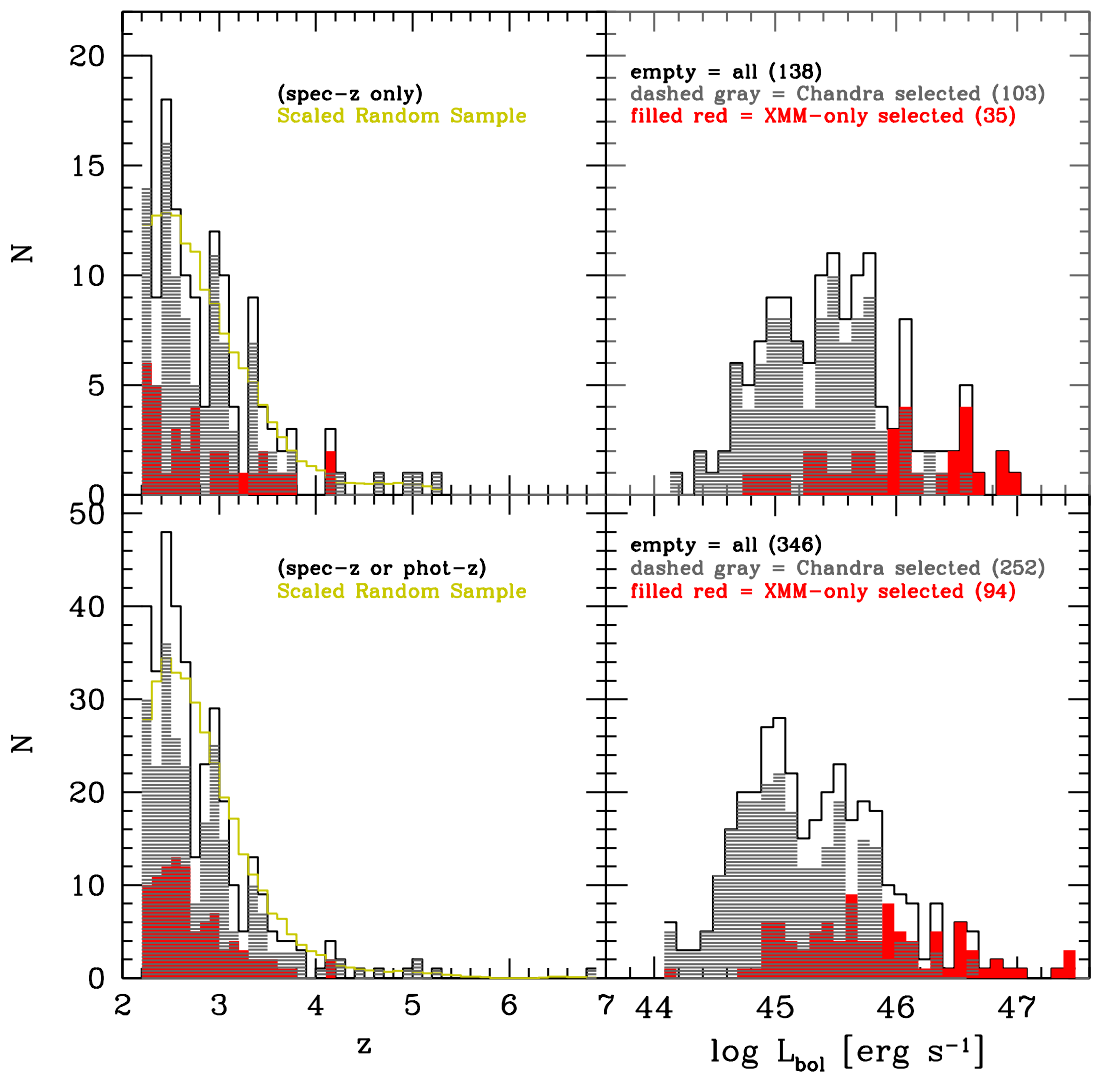}
\caption{\footnotesize Redshift and intrinsic bolometric luminosity distributions of \textit{Chandra}-COSMOS AGN (dashed gray histogram), XMM-only selected AGN (filled red histogram) and of the combined catalog (solid black histogram) of 138 COSMOS AGN with known spec-zs 
$\geqslant$2.2 (upper quadrants) and of 346 COSMOS AGN with known spec or phot-zs $\geqslant$2.2 (lower quadrants). The empty gold histograms in the left panels show the redshift distributions of the random catalogs obtained using a Gaussian smoothing with $\sigma$=0.2.}
\label{fig1}
\end{figure*}

\begin{figure*}
\plotone{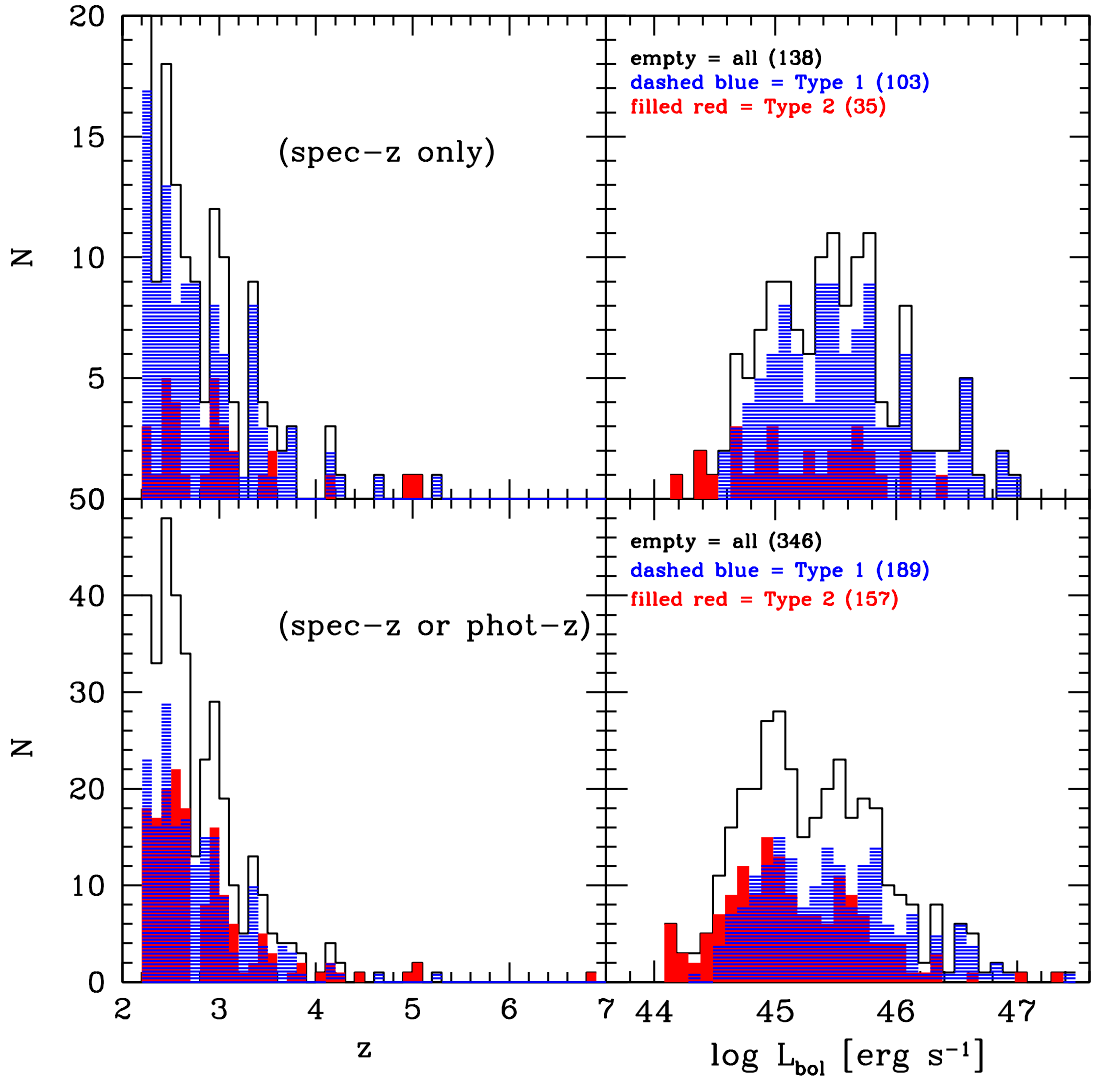}
\caption{\footnotesize Redshift and intrinsic bolometric luminosity distributions of Type 1 AGN (dashed blue histogram), Type 2 AGN (filled red histogram) and of the combined catalog (solid black histogram) of 138 COSMOS AGN with known spec-zs $\geqslant$2.2 (upper quadrants) and of 346 COSMOS AGN with known spec or phot-zs $\geqslant$2.2 (lower quadrants). The mean values of the distributions are quoted in Table 1.} 
\label{fig2}
\end{figure*}

\begin{figure*}
\plottwo{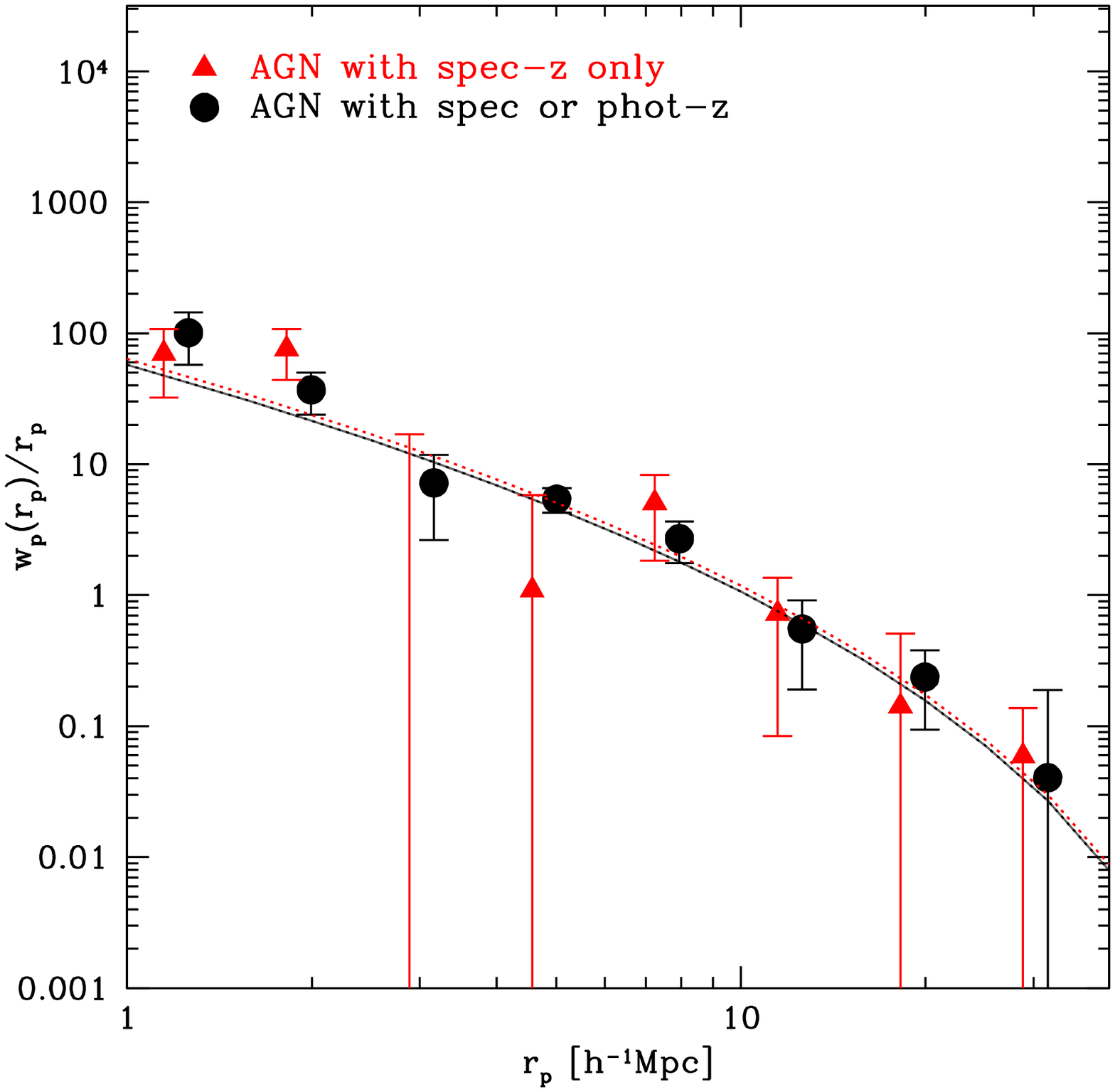}{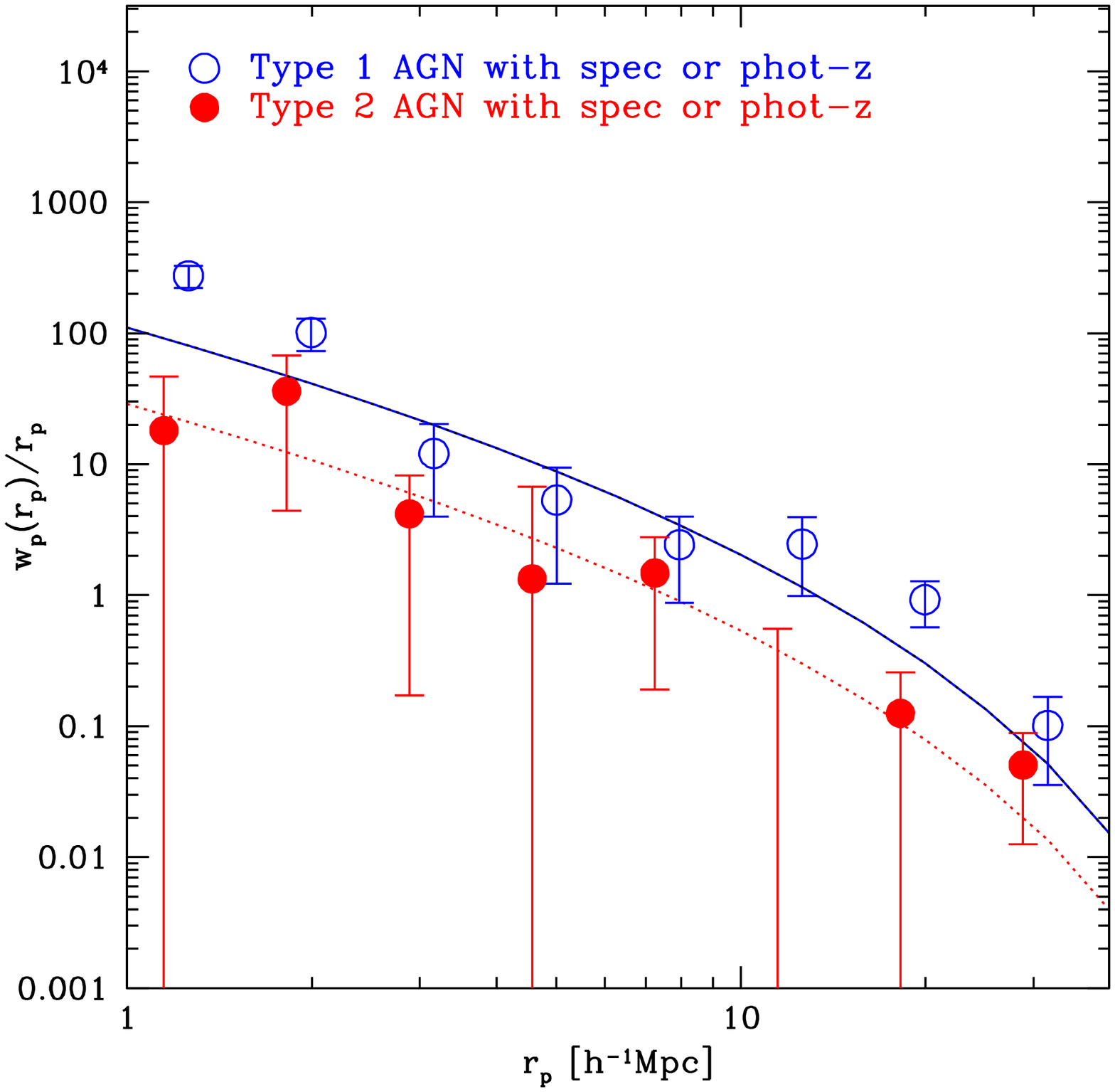}
\caption{\footnotesize  Left Panel: Projected 2PCF of 346 COSMOS AGN with phot or spec-zs, when available, $\gtrsim$ 2.2 (black circles) and 138 COSMOS AGN with known spec-zs $\gtrsim$ 2.2 (red triangles). The 1$\sigma$ errors on $w_p(r_p)$ are the square root of the diagonal components of the covariance matrix, which quantifies the level of correlation between different bins. Right panel: Projected 2PCF of 189 Type 1 (blue empty circles) and 157 Type 2 (red circles) COSMOS AGN with known spec or phot-zs $\gtrsim$ 2.2. The lines mark the AGN 2-halo term as defined in Equation 3, i.e.  
$b^2 w_{DM}^{2-h}(r_p)$, where $w_{DM}(r_p)$ is the DM 2-halo term evaluated at the mean redshift of the samples.} 
\label{fig3}
\end{figure*}


\section{AGN Catalog}\label{sec:AGNcat}

The \emph{Cosmic Evolution Survey} (COSMOS, Scoville et al. 2007) 
is a panchromatic photometric and spectroscopic 
survey of 2 deg$^2$ of the equatorial sky,
observed by the most advanced astronomical facilities,
with imaging data from X-ray to radio.
The inner part of the COSMOS field ($\sim 0.92$ deg$^2$) has been
imaged for a total of 1.8 Ms by \textit{Chandra}, while 
XMM-\emph{Newton} surveyed $2.13$ deg$^2$, 
for a total of $\sim$ 1.55 Ms.
Large samples of point-like X-ray sources
detected in the 0.5-10 keV energy band, are presented in the \textit{Chandra}-COSMOS
(C-COSMOS) point-like source catalog (1761 objects; Elvis et al 2009,
Civano et al. 2012) and in the XMM COSMOS multiwavelength catalog 
(1822 objects; Cappelluti et al. 2009, Brusa et al. 2010).
Of the 1822 XMM COSMOS sources, 945 have been detected by \textit{Chandra}.
Extensive spectroscopic campaigns have been carried out
in the field, providing a total of 890 and 1069 unique, good quality
spectroscopic redshifts (spec-zs) for XMM COSMOS and C-COSMOS sources,
respectively. In addition, photometric redshifts (phot-zs) 
for all the XMM COSMOS sources and for $\sim$96\% of the C-COSMOS sources
have been obtained exploiting the COSMOS multiwavelength
database and are presented in Salvato et al. (2009, 2011).

The prime interest of this paper is to investigate the clustering 
properties of X-ray AGN at z$\sim$3.
To this end, we use the catalog of C-COSMOS sources 
and we limit to a sample of 252  
AGN detected in the soft band, with phot or spec-zs, when 
available, $\geq$ 2.2.
In addition, we include in the analysis 94 AGN with spec or phot-zs
$\gtrsim$2.2, which are
outside the inner region observed by \textit{Chandra} 
and then XMM-only detected.
Then the final sample includes
a total of 346 COSMOS AGN.

The spectroscopic or photometric classification is
available for each AGN on the basis of 
a combined X-ray and optical classification
if spectra are available, or by the type of template that best
fits the photometry of the source.
We classify as Type 1 those sources with at least one broad
(FWHM $>$ 2000 km s$^{-1}$)
emission line in their spectra or fitted with
the template of an unobscured AGN. On the contrary, Type 2
AGN are defined as sources showing narrow emission lines or absorption lines
only, or well fitted by an obscured AGN template.
More details on this classification method
are presented in Brusa et al. (2010) and Civano et al. (2012).
Following the criteria described above, 189/346 and 
157/346 COSMOS AGN have been classified as
Type 1 and Type 2 , respectively.

The intrinsic 2-10 keV rest-frame luminosity, corrected for absorption,
is known for all the XMM COSMOS sources from the 
X-ray spectral analysis as described in
Mainieri et al. (2007) and Lanzuisi et al. (2013).  
Unfortunately, for C-COSMOS AGN the X-ray spectral 
analysis can be performed for only a few bright sources.
However, at z$>$2, the hardness ratio is a good measure of the 
presence of high column densities and absorption. 
For this reason, we use the hardness ratios known for all
the C-COSMOS sources to derive the absorption column density
and then the 
de-absorbed X-ray luminosities.
In detail, for each C-COSMOS AGN,
we derive the rest frame
luminosity from the soft flux, assuming a power-law model 
with $\Gamma$=1.8. The Galactic absorption is set to 
N$_{H,Gal}$=2.6$\times$10$^{20}$ cm$^{-2}$, i.e. the value 
in the direction of the COSMOS field (Kalberla 
et al. 2005). The rest frame luminosity is then corrected for the intrinsic absorption 
using the hardness ratio.

Finally, we derive the bolometric luminosities for the entire sample of
346 COSMOS AGN using the bolometric correction k$_{bol}$ 
quoted in Lusso et al. (2012, see Table 2). In this conversion, 
we properly take into account the different classification of the
sources (Type 1 or Type 2) and the band of the 
de-absorbed rest-frame X-ray luminosity.

As shown in Figure \ref{fig1}, the intrinsic bolometric luminosity
of the entire sample of 346 COSMOS AGN,
spans $\sim$ 3 order of magnitude, from 10$^{44}$ to $\sim$10$^{47}$
erg s$^{-1}$. The mean and the dispersion of the distribution are
(45.3, 0.6) in log and in unit of erg s$^{-1}$. Therefore, this sample is 
dominated by moderate luminosity X-ray AGN with a mean bolometric
luminosity $\sim$ 2 orders of magnitude lower than that of luminous optical 
quasars at similar redshift (Shen et al. 2009). 
Due to the lower limiting flux of \textit{Chandra} 
with respect to XMM detections, the distribution of bolometric luminosities 
of C-COSMOS AGN peaks at lower values.
The intrinsic L$_{bol}$ distributions of 189 Type 1 and
157 Type 2 COSMOS AGN are shown in Figure \ref{fig2}, 
with mean and dispersion equal to 
(45.47, 0.58) and (45.15, 0.58) for Type 1 and 2, respectively
(in log and units of erg s$^{-1}$).

In order to evaluate the effect of using photometric
redshifts in the clustering measurements, we also construct 
a sample of 138 COSMOS AGN detected in the soft band, 
with available spec-zs $\geq$ 2.2. The redshift and 
intrinsic bolometric luminosity distributions for this sample are shown in 
Figure \ref{fig1} and the corresponding mean values 
are quoted in Table 1. 
Following the spectroscopic classification, the sample 
has been divided onto 107 Type 1 and 31 Type 2 COSMOS AGN (see Figure \ref{fig2}).

\section{2PCF and AGN bias factor}
\label{sec:2pcf}

Measurement of the two-point correlation function (2PCF)
requires the construction of a random catalog with the same
selection criteria and observational effects as the real data.
To this end, we construct a random catalog 
where each simulated source is placed at a random position in the sky, with
flux randomly extracted from the catalog of real source fluxes
(e.g. Gilli et al. 2009, Allevato et al. 2011).
Following this method, the simulated source is kept in the random sample 
if its flux is above the sensitivity map value at
that position \citep{Miy07, Cap09}.
We prefer this method with respect to the one that keeps 
the angular coordinates unchanged as that approach has the disadvantage of removing 
the contribution to the signal due to angular clustering.
Nevertheless, Gilli et al. (2005, 2009) and Koutoulidis et al. (2013) have shown that 
there is only a small difference ($\sim$15\%) in the clustering signal
derived with the two different procedures.

The corresponding redshifts of the random 
objects are assigned based on the smoothed redshift distribution 
of the real AGN sample. Specifically, we assume a Gaussian 
smoothing length $\sigma_z = 0.2$. This is
a good compromise between scales 
that are too small, which would suffer from 
local density variations, and those that are too 
large, which would oversmooth the distribution. 
The redshift distribution of COSMOS AGN 
and of the random samples
are shown in Figure \ref{fig1}.

We estimate the projected 2PCF 
function $w_p(r_p)$ by using \citep{Dav83}:
\begin{eqnarray}\label{eq:integral}
w_{AGN}(r_p) = 2 \int_0^{\pi_{max}} \xi(r_p,\pi) d\pi 
\end{eqnarray}
where $ \xi(r_p,\pi)$ is defined in \citet[LS]{Lan93} as:
\begin{equation}\label{eq:LZ}
\xi = \frac{1}{RR} [DD-2DR+RR]
\end{equation}
The LS estimator is described as the ratio between 
AGN pairs in the data sample and those in the 
random catalog, as a function of the projected comoving 
separations between the objects (in the directions perpendicular, 
r$_p$ and parallel, $\pi$ to the line-of-sight).
The choice of $\pi_{max}$ is a compromise between having
an optimal signal-to-noise ratio and reducing the excess noise
from high separations. 
Usually, the optimum $\pi_{max}$ value can be determined by
estimating $w_p(r_p)$ for different values of $\pi_{max}$ and finding 
the value at which the 2PCF levels off. Following this
approach, we fixed $\pi_{max}$ = 100 h$^{-1}$Mpc in estimating the
2PCF of 346 COSMOS AGN with known spec or phot-zs. 
For the smaller sample of
138 COSMOS AGN with available spec-zs, we set
$\pi_{max}$ = 40 h$^{-1}$Mpc. The larger $\pi_{max}$ adopted
in the former case is due to the use of phot-zs.

In the halo model approach (e.g. Miyaji et al. 2011, Krumpe et al. 2012), 
the 2PCF is modelled as the 
sum of contributions from AGN pairs within individual DMHs 
(1-halo term, r$_p < 1$ Mpc h$^{-1}$) and in different DMHs 
(2-halo term, r$_p \gtrsim 1$ Mpc h$^{-1}$). The superposition of 
the two terms describes the shape of the observed 2PCF.
In this context, the bias parameter $b$ 
reflects the amplitude of the AGN 2-halo term relative to the underlying 
DM distribution, i.e.:
\begin{equation}\label{eq:b}
w_{mod}(r_p,z) = b^{2} w_{DM}^{2-h}(r_p,z)
\end{equation}
We first estimate the DM 2-halo term at the mean redshift of the sample, using:
\begin{equation}
w_{DM}^{2-h}(r_p)=r_p \int_{r_p}^{\infty} \frac{\xi^{2-h}_{DM}(r)rdr}{\sqrt{r^2-r_p^2}}
\end{equation}
where
\begin{equation}\label{eq:2-halo}
\xi^{2-h}_{DM}(r)=\frac{1}{2\pi^2}\int P^{2-h}(k)k^2 \left[ \frac{sin(kr)}{kr} \right]  dk
\end{equation}
$P^{2-h}(k)$ is the Fourier Transform of the linear power spectrum,
assuming a power spectrum shape parameter $\Gamma = 0.2$ which corresponds to
$h=0.7$.

\section{Results}
\label{sec:results}

\subsection{Bias Factors and DMH masses}

The projected 2PCF function $w_p(r_p)$ of 346 COSMOS AGN is 
shown in the left panel of Figure \ref{fig3}, in the range of $r_p$ = 
1-30 h$^{-1}$ Mpc. The 1$\sigma$ errors on $w_p(r_p)$ are the square root of the 
diagonal components of the covariance matrix (Miyaji et al. 2007, Krumpe et al. 2010), 
which quantifies the level of correlation between different bins.
Following Eq. \ref{eq:b}, we derive the best-fit bias by using a 
$\chi^2$ minimization technique with 1 free parameter,
where $\chi^2 = \Delta^T M^{-1}_{cov} \Delta$. In detail, 
$\Delta$ is a vector composed of $w_{AGN}(r_p)-w_{mod}(r_p)$ (see Equations \ref{eq:integral} and \ref{eq:b}), $\Delta^T$ is its transpose
and M$^{-1}_{cov}$ is the inverse of covariance matrix.
The latter is used in the fit to take into account the
correlation between errors.
We find that, at $\langle z \rangle$=2.8, 
COSMOS AGN have a bias of b=3.85$^{+0.21}_{-0.22}$, 
where the 1$\sigma$ errors correspond to 
$\Delta \chi^2$ = 1.

We then relate the large-scale bias to a typical mass
of the hosting halos, following the bias-mass relation 
$b(M_h, z)$ defined by the ellipsoidal collapse model of 
Shen et al. (2001) and the analytical approximation of van den Bosch (2002).
We find that COSMOS AGN at $\langle z \rangle$=2.8 inhabit 
DMHs with log M$_h$ = 12.37$^{+0.10}_{-0.09}$.

\begin{deluxetable}{cccccc}
\tabletypesize{\scriptsize}
\tablewidth{0pt}
\tablecaption{Properties of the AGN Samples \label{tbl-1}}
\tablehead{
\colhead{Sample} &
\colhead{N} & 
\colhead{$\langle z \rangle$} &
\colhead{log $\langle L_{BOL} \rangle$}  &
\colhead{$b$} &
\colhead{logM$_h$} \\
\colhead{} &
\colhead{z$>$2.2} &
\colhead{} &
\colhead{erg s$^{-1}$} &
\colhead{} &
\colhead{h$^{-1}$M$_{\odot}$} }
\startdata
\multicolumn{6}{c}{\textit{Only spec-zs}}\\
All AGN & 138  & 2.86  & 45.50 & 3.94$^{+0.45}_{-0.46}$ &12.36$^{+0.17}_{-0.21}$\\
Type 1 & 107 & 2.82 & 45.58 & 4.93$^{+0.55}_{-0.52}$ & 12.75$^{+0.15}_{-0.16}$ \\
Type 2 & 31 & 2.96 & 45.22 & ... & ...\\
\multicolumn{6}{c}{\textit{Spec or phot-z}}\\
All AGN & 346 & 2.8 & 45.32 & 3.85$^{+0.21}_{-0.22}$ & 12.37$^{+0.10}_{-0.09}$\\
Type 1 & 189 & 2.79  & 45.47 & 5.26$^{+0.35}_{-0.39}$ & 12.84$^{+0.10}_{-0.11}$ \\  
Type 2 & 157 & 2.81 & 45.15 & 2.69$^{+0.62}_{-0.69}$ & 11.73$^{+0.39}_{-0.45}$\\
\enddata

\end{deluxetable}

\begin{figure*}
\plottwo{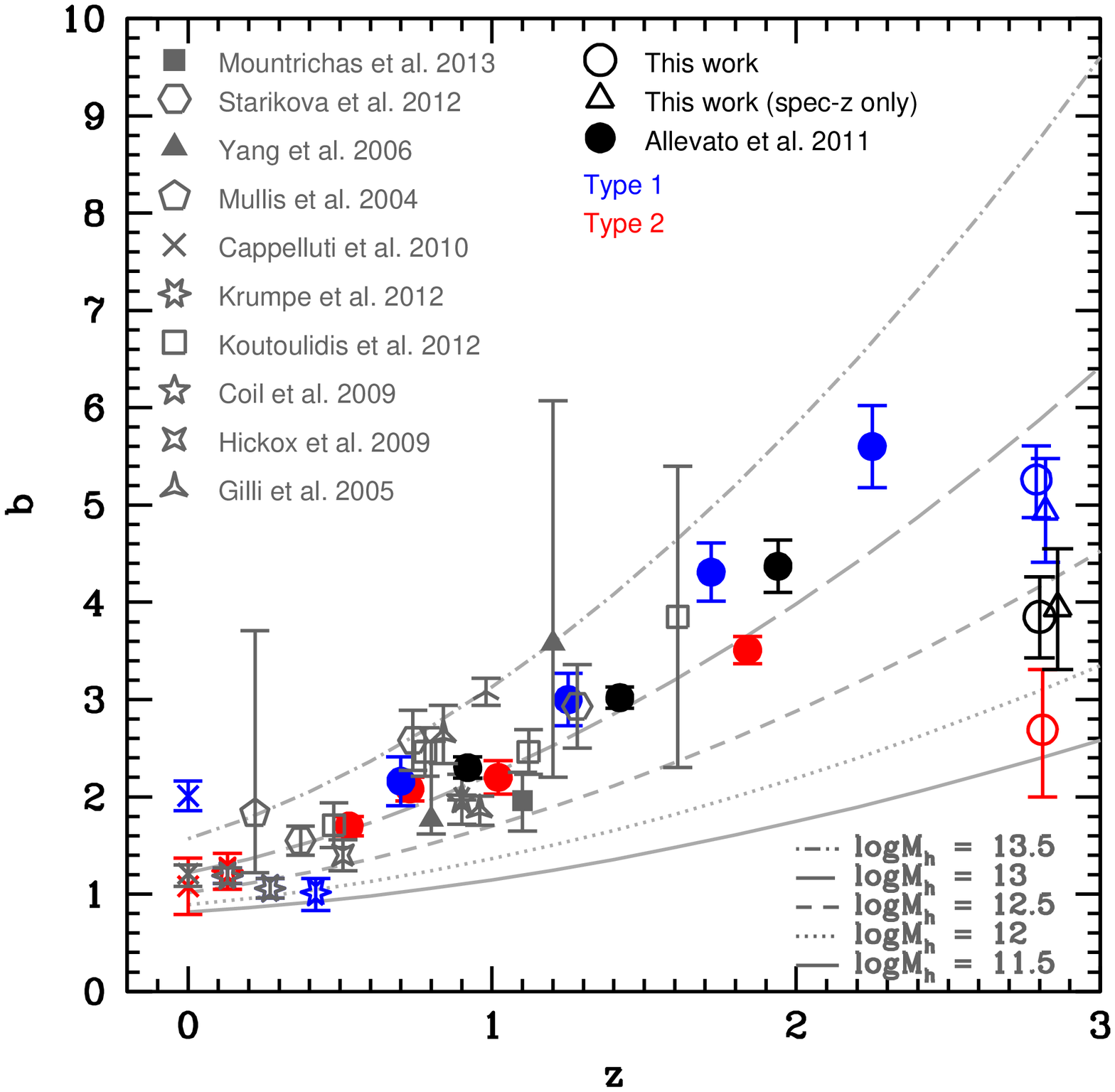}{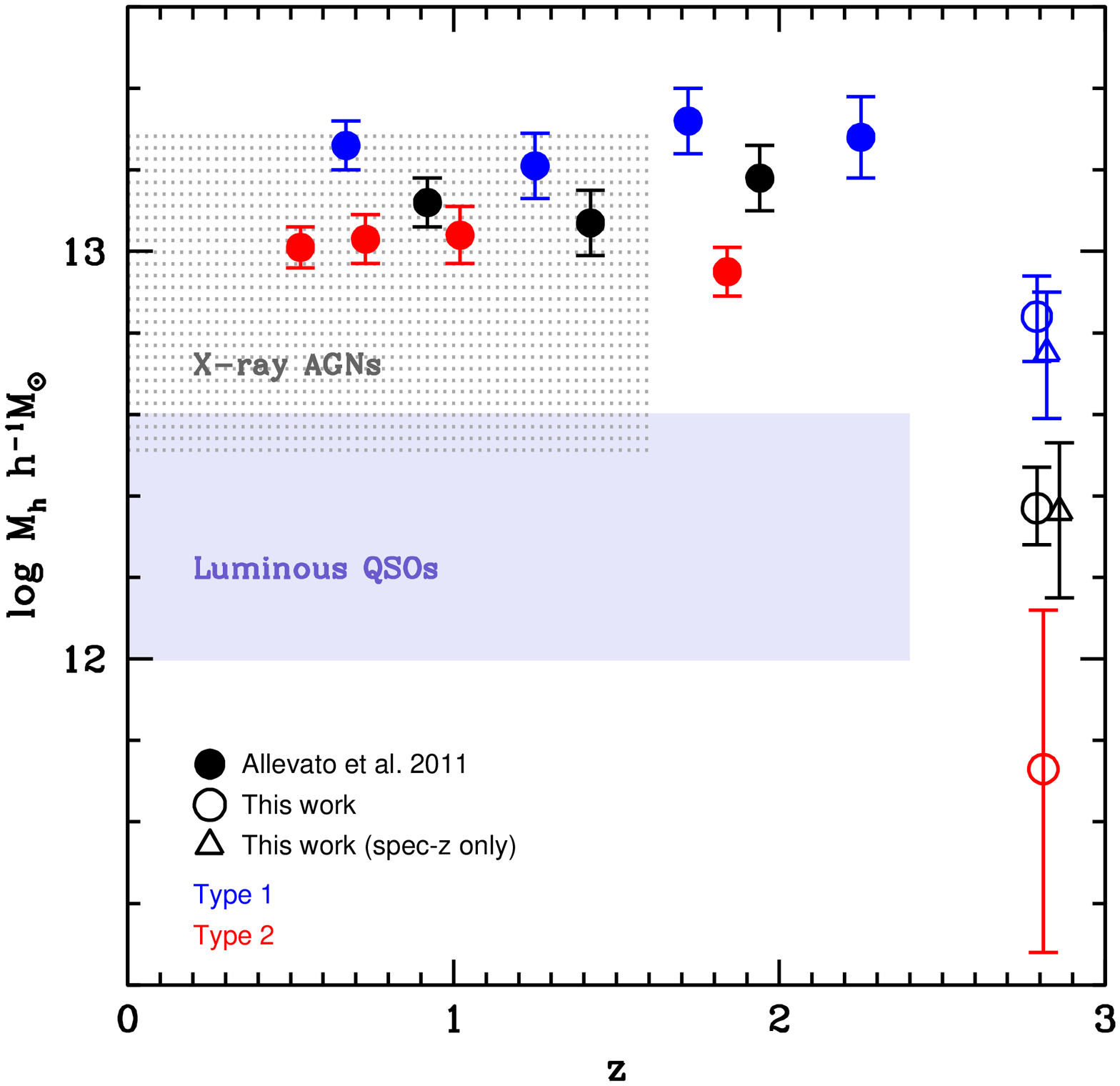}
\caption{\footnotesize Left Panel: Bias factor as a function of redshift for moderate luminosity X-ray selected (grey), Type 1 (blue) and Type 2 (red) AGN, as derived in previous works according to the legend. The empty triangles and circles at z$\sim$3 mark the bias factors of our sample of moderate luminosity COSMOS AGN as estimated in this work. For comparison, the filled circles show the bias derived in  Allevato et al. (2011) at z$\lesssim$2.2 for XMM COSMOS AGN with similar luminosities. The dotted lines underline the expected redshift evolution of the bias of DMHs with constant mass of 11.5, 12, 12.5, 13, 13.5 h$^{-1}$ M$_{\odot}$ in log scale and in unit of M$_{\odot}$h$^{-1}$ (from bottom to top). The bias-mass relation is based on the ellipsoidal collapse model of Sheth et al. (2001). Right Panel: Corresponding typical DMH mass of our sample of COSMOS AGN at z$\sim$3, divided in Type 1 (blue) and Type 2 (red) AGN,
estimated following the bias-mass relation b(Mh,z) described in van den Bosch (2002) and Sheth et al. (2001). For comparison, the filled circles mark the redshift evolution of the typical DMH mass of XMM COSMOS AGN at z$\lesssim$2.2, as derived in Allevato et al. (2011).
The halo mass range is shown also for optically selected luminous quasars (shaded region, Croom et al. 2005, Porciani et al. 2004, Myers et al. 2006, Shen et al. 2009, Ross et al. 2009, da Angela et al. 2008), and X-ray selected AGN (dotted shaded region, Gilli et al. 2005, Coil et al. 2009, Cappelluti et al. 2010, Allevato et al. 2011, Krumpe et al. 2010, 2012, Hickox et al. 2009, Mountricas et al. 2013, Koutoulidis et al. 2012, Starikova et al. 2012).}
\label{fig5}
\end{figure*}

Usually, phot-zs are characterized 
by large uncertainty. Hickox et al. (2012) showed that,
even uncertainties of $\sigma_z$=0.25(1+z)
cause the clustering amplitude of AGN
cross-correlated with galaxies in the Bootes field, to 
decrease by only $\sim $10 per cent.
COSMOS AGN has the advantage that, at z$>$2,
$\sigma_z/(1+z) <$ 0.05 (Salvato et al. 2011).
Hence
we do not expect a significant
difference from the 2PCF derived using only spec-zs.

In order to verify this, we measure the clustering 
signal for a smaller sample of 138 COSMOS AGN
with available spec-zs $\geq$2.2.
The left panel of Figure \ref{fig3} compares the projected 2PCFs
estimated with the two different AGN samples. As expected,
by using a larger AGN sample with both phot-zs and
spec-zs, we improve the statistics and so the quality of the signal.
However, we find consistent bias factors, 
irrespective of including photometric redshifts (see Table 1).
This result suggests that the use of a 
significant fraction of AGN with known phot-zs 
is not affecting the result systematically. Instead, we are 
improving the statistics, almost tripling the number of AGN.

We investigate whether Type 1 COSMOS
AGN are more strongly clustered than Type 2 objects, as 
already observed at low redshift (e.g. Cappelluti et al. 2010, Allevato et al. 2011).
The right panel of Fig. \ref{fig3} shows the projected
2PCF of Type 1 and Type 2 AGN with known phot and spec-zs $\geq$2.2.
We find that unobscured COSMOS AGN reside in 
more massive halos compared to obscured AGN.
In fact, we measure a best-fit bias  
equal to b$_{unob}$=5.26$^{+0.35}_{-0.39}$ for Type 1 
and b$_{ob}$=2.69$^{+0.62}_{-0.69}$
for Type 2 AGN, respectively. These bias
factors correspond to typical DMH masses of logM$_h$= 
12.84$^{+0.10}_{-0.11}$ and logM$_h$= 
11.73$^{+0.39}_{-0.45}$ h$^{-1}$M$_{\odot}$, respectively.

We check that the bias factor of 
Type 1 COSMOS AGN does not change
when limiting the analysis to a sample of 107 unobscured sources
with available spec-zs $\geq$2.2. Unfortunately, we cannot test
the effect of phot-zs in measuring the 2PCF of obscured
sources, given the small number of Type 2 objects (31) with 
spec-zs $\geq$2.2.
The best-fit bias factors and the corresponding
typical DMH masses for each subsample of COSMOS AGN
used in this work are shown in Table 1.

\subsection{Redshift Evolution of the AGN bias}

The left panel of Figure \ref{fig5} shows the bias factors 
derived for our sample of COSMOS AGN at z$\sim$3, along with
a collection of values estimated in previous studies at lower
redshifts.
The different lines mark the redshift evolution of the bias 
corresponding to different constant DMH masses,
as predicted by the ellipsoidal collapse model of Sheth et al. (2001).
The filled circles show 
the redshift evolution of 
the bias for a comparable sample of moderate luminosity XMM COSMOS AGN 
at z$\lesssim$2, as presented 
in Allevato et al. (2011).

The right panel of Figure \ref{fig5} shows the redshift evolution
of the corresponding typical DMH masses derived following 
the bias-mass relation defined by the ellipsoidal collapse 
model of Shen et al. (2001).
The general picture at z$\lesssim$2 is that the bias of 
moderate luminosity X-ray AGN increases with redshift 
tracing a constant group-sized halo mass.
Allevato et al. (2011) 
have shown that XMM COSMOS AGN (with L$_{bol} \sim 10^{45.2}$ erg s$^{-1}$) 
reside in DMHs with constant mass equal to logM$_h$ = 13.12$\pm$0.07 h$^{-1}$ M$_{\odot}$
up to z$\sim$2 (Figure \ref{fig5}, filled black points).

By contrast, at z$\sim$3, we found that our COSMOS AGN
with similar luminosities,
inhabit less massive DMHs (Figure \ref{fig5}, open black points)
with logM$_h$ = 12.37$^{+0.10}_{-0.09}$ h$^{-1}$M$_{\odot}$.
This result is significant at the 6.2$\sigma$ level.

A similar trend has been observed for both Type 1
and Type 2 COSMOS AGN. 
Allevato et al. (2011) found that unobscured AGN reside in 
slightly more massive halos than obscured AGN up to z$\sim$2.2 
(logM$_h$ = 13.28$\pm 0.07$ and 13.00$\pm 0.06$ h$^{-1}$M$_{\odot}$,
respectively).
Instead, our results at z$\sim$3 require, 
compared to z$\lesssim$2 results, a  
drop in the halo mass  
to logM$_h$ = 12.84$^{+0.10}_{-0.11}$ (3.6$\sigma$ result) and 
11.73$^{+0.39}_{-0.45}$ h$^{-1}$M$_{\odot}$ (3$\sigma$ result) for Type 1 and Type 2 
COSMOS AGN.


\section{Discussion}
\label{sec:disc}

In the following sections we will discuss our results
\begin{figure}
\plotone{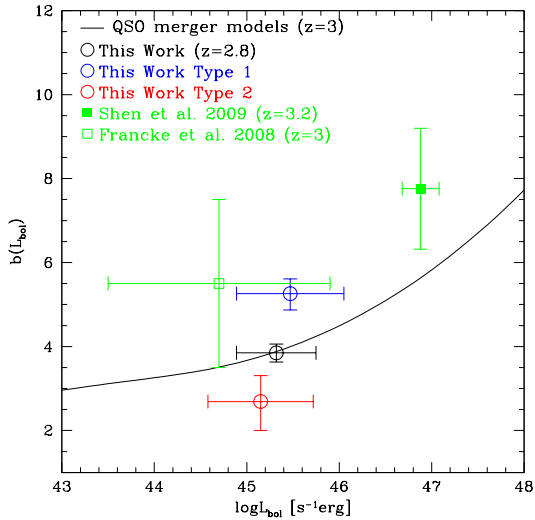}
\caption{\footnotesize Bias factor as a function of bolometric luminosity for optically selected SDSS quasars at z=3.2 (Shen et al. 2009, green squares), X-ray selected AGN in the ECDFS at z=3 (Francke et al. 2008, green open square) and our results according to the legend. For our data points, the errors on the L$_{bol}$ axis correspond to the dispersion of the bolometric luminosity distributions of each subset. The continuous line shows the luminosity evolution of the bias at z = 3, predicted with the theoretical model of Shen (2009), which assumes a quasar phase triggered by a major merger (see the text for more details).} 
\label{fig6}
\end{figure}
and we will attempt to answer the question why the
redshift evolution of the bias changes at z$\geqslant2.2$.

First of all we note that the number density of 10$^{13}$ h$^{-1}$M$_{\odot}$ 
mass halos tends to evolve
faster beyond z$\sim$3, with a progressive drop in the 
abundance of massive (and rarer)
host halos at high redshifts.
This fact alone suggests a possible increase with redshift in the ratio 
between the BH mass and host halo mass, with a mapping of 
moderate luminosity AGN in progressively less massive halos at higher redshift. Independent studies
support such a type of evolution (e.g. White, Martini \& Cohn 2008, 
Shankar et al. 2010).

\subsection{Major merger models at z $\sim$ 3}

In this section 
we try to explain our
clustering measurements at z$\sim$3 within hierarchical mergers scenario.

Figure \ref{fig6} shows a collection of bias estimates 
for broad-line optically selected SDSS
quasars at z=3.2 (Shen et al. 2009), 
X-ray selected AGN in the ECDFS at z=3 (Francke et al. 2008) 
and the results presented in this work,
as a function of L$_{bol}$. 
For our data points, the errors on the L$_{bol}$ axis correspond to the 
dispersion of the bolometric luminosity distributions for the 
different subsets. 
Our sample of COSMOS AGN 
and X-ray ECDFS AGN with slightly lower luminosities,
have consistent bias factors within 1$\sigma$.
However, our estimate has significantly smaller uncertainty,
given the larger number of sources used in the analysis.
On the contrary, Shen et al. (2009)
measured a slightly higher bias ($<$2$\sigma$) for  
luminous Type 1 quasars with bolometric luminosity $\sim$ 2 orders
of magnitude higher with respect to our sample of unobscured AGN.

The continuous line marks
the predicted bias as a function of bolometric luminosity, 
computed according to 
the framework of the growth and evolution of SMBHs 
presented in Shen (2009, see also Shankar 2010) 
at z = 3. Their model assumes that quasar activity 
is triggered by major mergers of host halos 
(e.g. Kauffmann \& Haehnelt 2000). In addition, they 
assume that the resulting AGN light curve follows a universal form 
with its peak luminosity correlated with the (post-)merger halo mass.

The major merger model was adapted to fully reproduce the optical/X-ray data 
at z$<$2-3 (Shankar et al. 2010, Allevato et al. 2011). 
With no additional fine-tuning the major merger model is quite successful in predicting the 
bias of COSMOS AGN at z$\sim$3 as a function of bolometric luminosity
and it is in broad agreement with the data points.
The drop in the typical DMH mass to few times 10$^{12}$ M$_{\odot}$h$^{-1}$ can then be explained assuming that, unlike z$\lesssim$2 XMM COSMOS AGN, COSMOS AGN at z$\sim$3 are triggered by galaxy major mergers.


In addition, the major merger model 
predicts a luminosity-dependent bias, with more luminous 
AGN inhabiting more massive DMHs. The evolution of the 
bias with the bolometric luminosity traced by the data
marginally confirms this trend. It is important to note
that at lower redshifts the bolometric luminosity dependence is 
significantly milder (e.g. Myers et al. 2006, Shen et al. 2009) or even 
reversed (e.g. Allevato et al. 2011), with moderate
luminosity AGN residing in more massive DMHs with respect to
luminous quasars. Such an evolutionary trend is difficult to reconcile 
with AGN triggering models, at different redshifts, based only on major mergers (Bournaud et al. 2011, Fanidakis et al. 2013, Di Matteo et al. 2011).

In the framework of the BH growth presented in Fanidakis et al. (2013, 2013a),
our results at z$\sim$ 3 can be interpreted   
in terms of absence of interplay between cold and hot-halo mode. 
In fact, they suggest the picture that, in the z$\sim$3-4 Universe,
the cold accretion mode (accretion during disk instabilities and galaxy mergers)
is solely responsible for determining the environment of moderate luminosity AGN,
while the AGN feedback is switched off.
Our results confirm that z$\sim$3 is the epoch when the hot-halo mode is
still a negligible fuelling channel. At lower redshifts both accretion modes
have to be taken into account. The hot halo mode becomes 
prominent only in DM haloes 
with masses greater than $\sim$ 10$^{12.5}$ h$^{-1}$M$_{\odot}$, where AGN feedback 
typically operates. 

\subsection{Fast growing BHs at z $\sim$ 3}

A major merger model can broadly reproduce the clustering of 
moderate luminosity AGN at z$\sim$3. This is because mergers are 
most efficient in halos of masses around $\sim$3$\times$10$^{12}$ M$_{\odot}$,
which is the typical mass scale inferred from our direct clustering 
measurements. However, this is not a proof of uniqueness of 
merger models as an explanation of our results.
Major galaxy mergers are not a requirement for efficient
fuel supply and growth, particularly for the earliest BHs. 
Alternatively, an early phase of fast BH growth could be induced 
by cosmic cold flows (e.g. Dekel et al. 2009, Di Matteo et al. 2012, Dubois et al. 2012) or
disk instabilities (e.g. Bournaud et al. 2012).
Cold flows and disk instabilities in high redshift disk galaxies 
operate on short time scales (unlike secular processes in low-z disks).
In addition, they are efficient, producing a mass inflow similar to a major merger but
spread over a longer period (then the duty cycle is higher).
%

\begin{figure}
\plotone{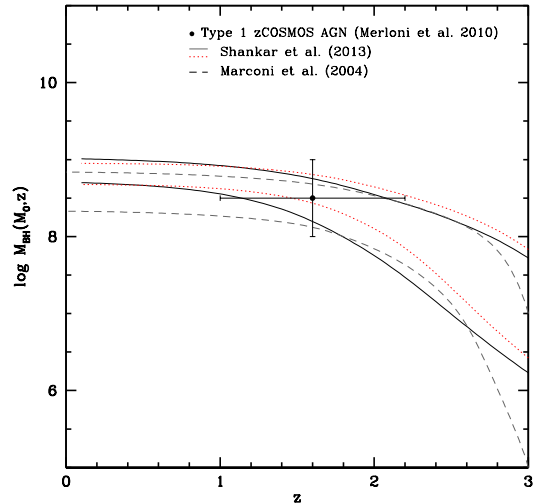}
\caption{\footnotesize Average accretion history of BHs of different mass M$_0$ at z = 0 as a function of redshift. The black continuous (red dotted) line marks a model with a redshift-independent (dependent) Gaussian distribution of the Eddington ratio with a fixed radiative efficiency (Shankar et al. 2013).  The black dashed line shows the growth history of BHs described in Marconi et al. (2004) with given starting mass at z = 3 computed using the Ueda et al. (2003) luminosity function and fixed duty cycle and radiative efficiency. The data point shows Type 1 zCOSMOS AGN in the redshift range 1$<$z$<2.2$ presented in Merloni et al. (2010) with BH masses in the range M$_{BH}$=10$^{8-9}$ M$_{\odot}$ and Eddington ratios $\lambda$=0.01-1.} 
\label{fig7}
\end{figure}

Different BH accretion models (e.g. Marconi et al. 2004,
Merloni et al. 2008, Shankar et al. 2004,2009,2013, Bournaud et al. 2011, Fanidakis et al. 2013, Di Matteo et al. 2011) broadly agree that 
$z \gtrsim 2$ is the epoch of rapid growth for both low and high-mass BHs.
This conclusion holds irrespective of uncertainties in 
duty cycle, AGN luminosity functions, or even input Eddington 
ratio distributions. Figure \ref{fig7} shows 
the average accretion history of different BH masses as described in 
Marconi et al. (2004) and Shankar et al. (2013). 

Due to the flux and volume limits of our
survey, our sample of Type 1 COSMOS AGN at z$\sim$3 mainly includes moderate
luminosity sources with L$_{bol} \sim 10^{45}$ erg s$^{-1}$.
Thus, it excludes low-luminosity AGN with typical BH
mass $\lesssim10^{6-7}$M$_{\odot}$,
or even bright quasars associated to
very massive BHs ($>10^{8-9}$M$_{\odot}$)
and L$_{bol}>10^{46}$ erg s$^{-1}$. This means that 
our sample of Type 1 COSMOS AGN is possibly representative
of AGN corresponding to raising population of fast (i.e. Eddington limited)
growing BHs with masses of $\sim$10$^{7-8}$ M$_{\odot}$
at z$\sim$3.

According to the BH accretion histories shown in Fig \ref{fig7}, 
these fast growing $\sim$10$^{7-8}$ M$_{\odot}$ mass BHs
evolve in BHs with mass of the
order of $\sim$10$^{8.5-9}$ M$_{\odot}$ at z=0.
This picture is consistent with the fact that
moderate luminosity Type 1 AGN in zCOSMOS at
z=1-2.2 (black circle in Figure  \ref{fig7}) have BH masses in the range
M$_{BH}$=10$^{8-9}$ M$_{\odot}$ with
Eddington ratios $\lambda$=0.01-1,
as shown in Merloni et al. (2010).

Following our results, we can argue that these
fast growing BHs at z$\sim$3 reside in DMHs with typical
mass of $\sim10^{12.8}$ h$^{-1}$M$_{\odot}$,
which is the mass inferred for Type 1 COSMOS AGN hosting halos.
Between z$\sim$3 and z$\sim$2 the typical halo mass 
of Type 1 COSMOS AGN increases. The 
BH accretion models suggest that these AGN are  
also rapidly growing their BH mass. The
duty cycle and Eddington
ratio, which are close to unity at z$\sim$3,
then decline with decreasing redshift.
This leads to a strong evolution of the number 
density of X-ray AGN at z$\gtrsim$3 as observed in COSMOS 
(Brusa et al. 2010, Civano et al. 2011), albeit for 
an AGN sample including both Type 1 and 2 objects.


The picture is completely different at z$<$2, 
where the bias of Type 1 AGN
starts to follow the constant DMH mass track.
The growth of BHs becomes more sub-Eddington (e.g., 
Vittorini et al. 2005, Shankar et al. 2013, and references 
therein) and their mass saturates to a constant value 
down to $z=0$ (Figure \ref{fig7}).
This flat host halo mass at z$\lesssim$2 for 
X-ray AGN with moderate luminosity might be due to high mass halos
switching to radio-loud, X-ray weak ($<10^{44}$ erg s$^{-1}$) AGN.
This limits the X-ray AGN population with moderate luminosity 
to halos with masses $\lesssim10^{13}$ M$_{\odot}$.
A plausible mechanism is that AGN feedback prevents 
gas from cooling in very massive halos. These radiatively 
inefficient, jet-dominated outbursts may be fueled by accretion 
directly from the hot gas halo and so are only possible in massive galaxies 
with large hosting halos 
(e.g. Fanidakis et al. 2011, 2012, 2013). 

The picture described above for Type 1 COSMOS AGN representative, 
of new fast-growing BHs at z$\sim$3, may or may not be valid for
Type 2 COSMOS AGN. In fact, for the latter, we do not know the
typical mass of the central BHs at any redshift.
However, it is not unreasonable to assume that for Type 2 AGN,
the observed redshift evolution of the bias reflects
lower massive BHs and DMH masses with respect to Type 1 AGN.
Lower massive halos are more abundant at z$\sim$3 and are 
characterized by a slower redshift evolution of the number density. 
The increase of the fraction of obscured AGN with redshift 
(e.g. Hasinger 2008, Merloni et al. 2014) supports 
this scenario in which Type 1 and Type 2 objects follow
different DMH mass tracks as a function of redshift.

\subsection{Type 1 vs Type 2 AGN}

In this section we discuss the clustering 
properties of 189 Type 1 and 157 Type 2 
COSMOS AGN with phot or spec-zs, when available, $\gtrsim$ 2.2.

We find a strong indication that unobscured AGN reside 
in 10 times more massive halos (see Table 1) with respect to obscured sources (3$\sigma$ result).
A difference in 
clustering between obscured and unobscured 
quasars rules out the simplest unified models (e.g. Urry \& Padovani 1995) 
in which obscuration is purely an orientation effect.

Type 1 and Type 2 COSMOS AGN 
have slightly different luminosities.
However, the difference in the bias factors between
Type 1 and Type 2 AGN can not be explained in terms of 
luminosity-dependent bias predicted by major merger models.
The curve in 
Figure \ref{fig6} predicts a milder
change of bias with luminosity in the range logL$_{bol} = 45.1-45.5$ 
erg s$^{-1}$.

This result would extend to z$\sim$3 that
found for z$<$2 Type 1 and Type 2 XMM COSMOS AGN
with similar luminosities (Allevato et al. 2011).

\section{Conclusions}

We use a sample of 346 moderate luminosity 
($\langle L_{bol} \rangle = 10^{45.3}$ erg s$^{-1}$)
COSMOS AGN based on \textit{Chandra} and \textit{XMM-Newton} data,
with known spec or phot-zs in the range 2.2$<$z$<$6.8.
Our main goal is to measure clustering amplitudes and 
to estimate characteristic DM halo masses at z $\sim$ 3.
We also obtain, for the first time at z $\sim$ 3, a highly significant 
clustering signal for Type 1 and Type 2 
COSMOS AGN.
This redshift range has never been used before
to investigate the clustering of obscured and unobscured AGN, 
at these luminosities. 
We model the 2PCF of COSMOS AGN with the halo model, 
which relates the large-scale bias to the amplitude
of the AGN 2-halo term relative to the underlying DM
distribution. We translate the
bias factor into a typical mass of the hosting
halos, following the bias-mass relation 
defined by the ellipsoidal collapse model of Sheth et al. (2001).
Key results can be summarized as follows.

\begin{enumerate}
 \item At z$\sim$3 Type 1 and Type 2 COSMOS AGN inhabit DMHs with typical 
 mass of logM$_h$ = 12.84$^{+0.10}_{-0.11}$ and 
 11.73$^{+0.39}_{-0.45}$ h$^{-1}$M$_{\odot}$. This result requires a 
 drop in the halo masses at z$\sim$3 compared to z$\lesssim$2 
 XMM COSMOS AGN with similar luminosities.
 \item At z$\sim$3 Type 1 COSMOS AGN reside in $\sim$10 times more massive halos
 compared to Type 2 COSMOS AGN, at 2.6$\sigma$ level. This result extends 
 to z$\sim$3 that found in COSMOS at z$\lesssim$2, and rules out the picture in 
which obscuration is purely an orientation effect.
 \item A plausible explanation of the drop in the halo 
 mass of COSMOS AGN might be that these moderate luminosity sources at z $\sim$ 3 are triggered by galaxy major mergers. In fact, major merger models are quite successful in predicting the halo mass
of COSMOS AGN and luminous SDSS quasars at z $\sim$ 3,
with the latter inhabiting more massive halos with respect to
moderate luminosity AGN.
\item Alternatively, we can argue that, at least for Type 1 COSMOS AGN, they are possibly representative of moderate luminosity AGN associated to an early phase of fast (i.e. Eddington limited) BH growth
induced by, for instance, cosmic cold flows or disk instabilities.
According with BH accretion models, these new fast 
growing BHs have  
masses of $\sim 10^{7-8}$ M$_{\odot}$ at z$\sim$3 which might evolve into 
$\sim 10^{8.5-9}$ M$_{\odot}$ mass BHs at z=0.
 \item Following our clustering measurements, we argue that 
 this fast BH growth at z$\sim$3, in Type 1 AGN with moderate luminosity, occurs in DMHs with typical 
mass of $\sim$ 6$\times 10^{12}$ h$^{-1}$M$_{\odot}$.
\end{enumerate}

Improving our understanding of the AGN triggering mechanisms 
at z $\sim$ 3 and beyond using AGN clustering measurements 
requires larger dataset. 
The COSMOS Legacy survey
(Civano et al. 2014, submitted)
with 1.45 deg$^2$ at 2$\times$10$^{-16}$ erg/cm$^2$/s
will provide the largest survey at this
depth ever performed. This will let us constrain
the faint end of the AGN luminosity function and 
of the BH mass functions at z$>$3. This regime, which is not otherwise
sampled, will allow us to understand the BH growth in the early
universe and to study the clustering properties of $\sim$ 350
expected luminous quasars and L$^{*}$ AGN at 3$<$z$<$6.

\acknowledgments
We thank the referee for a very helpful report.
We gratefully acknowledge the contributions of the entire COSMOS collaboration consisting of more than 100 scientists. More information on the COSMOS survey is available at http://www.astro.caltech.edu/∼COSMOS. 
VA and AF wish to acknowledge Finnish Academy award, decision 266918.
FC aknowledges the support of NASA contract 11-ADAP11- 0218. 
FS acknowledges partial support from a Marie Curie grant.
TM acknowledges supports from UNAM-PAPIIT 104113 and CONACyT Grant 179662.
We thank Alessandro Marconi for providing the tracks shown in Figure 5 and John Regan for helpful discussions.

\clearpage

\end{document}